\documentclass[prb,twocolumn,showpacs,amsmath,amssymb,superscriptaddress]{revtex4}
\usepackage{epsfig,epstopdf} % Transform eps to pdf
\usepackage{graphicx}% Include figure files
\usepackage{dcolumn}% Align table columns on decimal point
\usepackage{bm}
\usepackage{bbm}
\usepackage{ulem}
\usepackage{color}
\usepackage{slashed}
\usepackage{hyperref}
\usepackage{mathrsfs}
\usepackage{subfigure}
\usepackage{verbatim}
\usepackage{dsfont}
\usepackage{float}
\newcommand{\nc}{\newcommand}

\nc{\be}{\begin{equation}} \nc{\ee}{\end{equation}}
\nc{\bea}{\begin{eqnarray}} \nc{\eea}{\end{eqnarray}}
\nc{\bean}{\begin{eqnarray*}} \nc{\eean}{\end{eqnarray*}}
\nc{\dg}{\dagger} \nc{\ua}{\uparrow} \nc{\da}{\downarrow}

\begin{document}

\title{Interacting topological phases in thin films of topological mirror Kondo insulators}

\author{Rui-Xing Zhang}
\affiliation{Department of Physics, The Pennsylvania State
University, University Park, Pennsylvania 16802}

\author{Cenke Xu}
\affiliation{Department of Physics, University of California,
Santa Barbara, CA 93106}

\author{Chao-Xing Liu}
\affiliation{Department of Physics, The Pennsylvania State
University, University Park, Pennsylvania 16802}

\begin{abstract}
%Topological insulator has an insulating gap in the bulk while
%gapless excitations at the boundary.
%It was theoretically predicted that interaction can drastically
%change the classification of some topological
%insulators/superconductors, but this change due to interaction has
%not yet been verified in experiments. In this work,
We study the interaction effects on thin films of topological
mirror Kondo insulators (TMKI), where the strong interaction is
expected to play an important role. Our study has led to the
following results: (1) We identify a rich phase diagram of non-interacting TMKI
with different mirror Chern numbers in the monolayer and bilayer
thin films; (2) We obtain the
phase diagram with interaction and identify the regimes of
interaction parameters to mimic bosonic symmetry protected
topological phases with either gapless bosonic modes or
spontaneous mirror symmetry breaking at the boundary; (3) For the
spontaneous mirror symmetry breaking boundary, we also study
various domain-wall defects between different mirror symmetry
breaking order parameters at the boundary. Our results reveal that
the thin film TMKI serves as an intriguing platform for the
experimental studies of interacting topological phases.
\end{abstract}

\date{\today}
\maketitle

\section{Introduction}

A topological state is usually characterized by certain type of
topological invariant. As a consequence of their ``nontrivial
topology", topological states possess gapless modes at the
boundary of the sample~\cite{wen1995}. Recently, intense research
interests have been focused on the role of symmetry in the
classification of topological states and it was shown that
symmetries substantially enrich the family of topological states.
These new topological states, known as symmetry protected
topological (SPT) states~\cite{chen2012,chen2013}, have been
proposed for different types of symmetries in electronic systems,
including time-reversal invariant topological insulators
(TIs)~\cite{fu2007,zhang2009,hasan2010,qi2011}, topological
superconductors~\cite{qi2011}, topological crystalline
insulators~\cite{fu2011,hsieh2012} and
superconductors~\cite{shiozaki2014}. With the help of the
state-of-the-art first principles calculations, these theoretical
predictions have successfully led to the experimental discovery of
different types of SPT phases in electronic systems. Examples
include two dimensional (2D) TIs in HgTe/CdTe quantum wells
\cite{bernevig2006,konig2007}, InAs/GaSb quantum wells
\cite{liu2008,knez2011}, {\it et al}, three dimensional (3D) TIs
in Sb/Bi alloy, Bi or Sb-based chalcogenides
\cite{zhang2009,chen2009,xia2009,hsieh2009}, {\it et al}. More
recently, it has been proposed that SPT phases can also be
realized in photonic systems \cite{rechtsman2013,khanikaev2013}.

%The concept of topological matters has attracted great attention and gained wide interest in condensed matter physics. Symmetry usually plays an essential role of stabilizing bulk topology, which gives rise to a wide class of topological phases dubbed symmetry protected topological (SPT) phase. A SPT phase is characterized by an integer bulk topological invariant as well as its gapless boundary modes. The first known example of SPT phases is time reversal (TR) invariant topological insulators (TI) which have been experimentally confirmed in various materials. Among these TIs, topological Kondo insulators (TKI) have been theoretically proposed and experimentally observed in SmB$_6$. The uniqueness of TKI lies in its strong correlation nature, offering us the first example of interacting topological insulators. What is more, recent studies confirm the topological crystalline phases in SmB$_6$ characterized by a set of non-trivial mirror Chern numbers. This indicates that SmB$_6$ simultaneously falls into two different topological classes: $\mathbb{Z}_2$ class protected by TR symmetry and $\mathbb{Z}$ class protected by mirror symmetry.

Theoretically, it has been shown that strong interaction can
significantly change the classification of topological
insulators/superconductors. The first example is the one
dimensional (1D) interacting topological superconductor. For free
fermions, 1D topological superconductors with time-reversal
symmetry $\mathcal{T}$ and $\mathcal{T}^2 = +1$ (BDI class)
\cite{schnyder2008} has a $\mathbb{Z}$ classification (which is
characterized by an arbitrary integer number of Majorana fermion zero
modes at its boundary), but Fidkowski and Kitaev
\cite{fidkowski2010,fidkowski2011} pointed out that appropriate
$\mathcal{T}-$invariant interactions can render the ground state
of eight Majorana fermion zero modes gapped and nondegenerate,
which implies that the classification of these TSCs is reduced
from $\mathbb{Z}$ to $\mathbb{Z}_8$ under interaction. Examples
for such ``interaction reduced classification" in higher
dimensions were also
found~\cite{qiz8,yaoz8,zhangz8,levinguz8,chenhe3B,senthilhe3,xu16,youinversion}.
%More examples have been found in interacting bosonic systems.
%({\bf Cenke, could you please add one or two sentences to
%summarize the progress of interacting bosonic SPT phase?  })
%Despite a large number of theoretical work, there are still no
%experiments for strongly interacting SPT states due to the lack of
%appropriate material proposals, in sharp contrast to the rapid
%experimental development of SPT states in free fermion systems.

Unlike their fermionic analogues, strong interaction is demanded
to realize bosonic SPT (BSPT) states. However, so far theoretical
studies for BSPT states have been focused on theoretical
classification and field theory
descriptions~\cite{chen2012,chen2013,senthilashvin,luashvin,xuclass,wangguwen},
but most of the lattice models proposed for BSPT states in two and
three dimensions are usually complex and unrealistic. Very
recently, it was proposed that interacting BSPT phases can be
realized by introducing interactions to a 2D quantum spin Hall
system with two channels of helical edge states~\cite{bi2016} and
total spin $S^z$ conservation. These BSPT states which originate
from fermionic systems not only have gapless bosonic edge modes,
but also are separated from the trivial state with a purely
``bosonic quantum phase transition", which has been observed
numerically~\cite{Kevin2015,YYHe2016a}. It was also shown
theoretically that the bilayer graphene under both a strong out-of-plane magnetic field and Coulomb interaction mimics much of
the physics of a BSPT state~\cite{bi2016}. A natural question is whether similar physics can exist in other condensed matter systems.
%These works provide us with a
%feasible approach to study bosonic and fermonic SPT phases in
%interacting fermionic systems.

%Since then, a lot of work have been done in this direction and a recent work by Isobe and Fu shows that two dimensional mirror TCIs yield a classification reduction from $\mathbb{Z}$ to $\mathbb{Z}_4$. Ideally, one would expect that a 2D TCI with a large mirror Chern number $C_m$ ($C_m\geq 4$) is able to verify this classification reduction. However, most of the proposed 2D TCIs only carry a mirror Chern number of 2.

%Another promising direction is to construct bosonic SPT phases based on a interacting fermionic $\mathbb{Z}$ classified SPT system. In an earlier work, we have shown that in a bilayer graphene based quantum spin Hall (QSH) system, $U(1)_c$ charge symmetry, together with $U(1)_s$ spin symmetry, protects a BSPT phase. With $S_z$ conservation, this QSH insulator has a spin Chern number of 2. Noticing the similarities between a QSH insulator and a 2D TCI system, it is natural to ask whether similar physics can appear in a 2D interacting TCI system.

In this work, we explore interacting topological phases, in particular the possibility of BSPT state, in topological Kondo insulators~\cite{dzero2010}, which is a class of
topological insulator materials with strong interactions. In Kondo
insulators, a small band gap opens up due to the hybridization
between localized f-electrons and conducting d-electrons. It turns
out that this hybridization gap is topologically non-trivial,
making this class of materials also time-reversal invariant
TIs. Topological nature of this class of
materials,
including SmB$_6$ and YbB$_6$, has recently been experimentally confirmed by Ref. [\onlinecite{neupane2013,jiang2013,xu2013,xia2014,xu2014,neupane2015}]. %(\textcolor{red}{Here it is interesting that in YbB$_6$, people claim the surface states are non-Kondo, which means the band inversion is between p and d electrons...})
More recent studies reveal that mirror symmetry also plays a role
in some Kondo insulators, thus we will refer to these systems as
topological mirror Kondo insulators (TMKI). For example, it is
found that the topological nature of SmB$_6$ can either be
protected by time-reversal symmetry ($\mathbb{Z}_2$ class)
\cite{ye2013} or by mirror symmetry ($\mathbb{Z}$ class without
interaction) \cite{hsieh2012}.
%simultaneously falls into two different topological classes : $\mathbb{Z}_2$ class protected by TR symmetry and $\mathbb{Z}$ class protected by mirror symmetry \cite{hsieh2012}.
Due to the existence of f-electrons near the Fermi energy,
interaction is quite strong in Kondo insulators. However, so far
the studies on topological phases in TMKIs have essentially
followed the paradigm of free electron TIs.
%for current theories of topological phases in TKIs, interaction effect only affects electronic band structure, such as band gap, of this system, and the corresponding topological nature is still discussed in the context of topological classification for free fermions.
Therefore, it is natural to ask if interacting SPT phases,
especially BSPT states that are qualitatively beyond the the free
fermion limit, can be realized in TMKIs.

%Among these TIs, topological Kondo insulators (TKI) have been theoretically proposed and experimentally observed in SmB$_6$. The uniqueness of TKI lies in its strong correlation nature, offering us the first example of interacting topological insulators. What is more, recent studies confirm the topological crystalline phases in SmB$_6$ characterized by a set of non-trivial mirror Chern numbers. This indicates that SmB$_6$ simultaneously falls into two different topological classes: $\mathbb{Z}_2$ class protected by TR symmetry and $\mathbb{Z}$ class protected by mirror symmetry.

To address this question, we consider thin films of TMKIs, such as
SmB$_6$, and study topological phases of this 2D system. We assume
that the thin film is grown along the direction with mirror
symmetry, in which mirror Chern number\cite{hsieh2012}, an integer
topological invariant defined based on mirror symmetry, can be
utilized to characterize the topological nature of this system. By
tuning the thickness and hybridization parameters between different
orbitals of the thin films, we find a rich phase diagram, which
stems from the competition between strong hybridization effect and
quantum confinement effect (QCE). Topological phases with
different mirror Chern numbers (as large as 6) can be realized in
the phase diagram. Furthermore, we show that in addition to the
charge $U(1)$ symmetry and mirror symmetry, another effective
$U(1)$ symmetry can be defined, which operates oppositely in two
mirror parity subspaces and is dubbed the ``pseudo-spin" $U(1)_m$
symmetry in this paper, at the single particle level.
%Different from the bulk TKI, the topology of a bilayer TKI thin film are found to be determined by both strong hybridization effect and additional quantum confinement effect (QCE). The competition between QCE and hybridization effect results in a colorful topological phase diagram with mirror Chern numbers to be as large as 6. The natural strong correlation in SmB$_6$ thin film makes it a very promising platform to test interaction induced classification reduction, as well as possible BSPT phases.
Based on the 2D model, we further study the interaction effect on
the topological mirror insulator phase with two copies of helical
edge states (mirror Chern number being 2) based on the Abelian
bosonization formalism. We notice that the ``pseudo-spin" $U(1)_m$
symmetry and mirror symmetry are playing different roles in the
interacting system. If interaction preserves the ``pseudo-spin"
$U(1)_m$ symmetry, the system can be driven into the BSPT phase
with only one gapless bosonic edge mode and hence central charge
$c=1$; if interaction breaks the ``pseudo-spin" $U(1)_m$ symmetry
while preserving the mirror symmetry, the edge states will be
gapped out due to spontaneous breaking of the mirror symmetry when
the interaction is relevant.
Depending on the form of interaction, different types of
interacting phases are discussed and the conditions for the
interacting BSPT phase are identified.

 % In the absence of translation symmetry, $U(1)_m$ breaking terms (Umklapp terms) naturally arise from Coulomb exchange interactions, while mirror symmetry is preserved. When these Umklapp terms are relevant, Mermin-Wagner theorem tells us that discrete mirror symmetry can be spontaneously broken, giving no protection to boson modes. This is a key difference between a interacting QSH insulator and a 2D interacting mirror TCI. We also discuss conditions where Umklapp terms are irrelevant and suppressed. Therefore, in this RG limit, the restored $U(1)_m$ symmetry and $U(1)_c$ symmetry together protect a BSPT phase, similar to the BSPT proposed in Ref. \cite{bi2016}.

\section{Two dimensional TCI phases in TMKI thin films}

\subsection{Model of TMKI thin films}

We start from a description of the 2D model for the TMKI thin
films. Following Ref. [\onlinecite{legner2015,baruselli2015}]
where various tight-binding models of SmB$_6$ have been summarized
and discussed, we consider a four band model described in Ref.
[\onlinecite{legner2014,legner2015}] due to its simplicity. This
model has been shown to reproduce energy dispersion from more
complicated multi-band models, as well as the first principles
calculations \cite{legner2015}. We notice that the obtained spin
texture from this model is not exactly the same as that from more
sophisticated models \cite{legner2015}, but this difference is not
essential for the physical mechanism discussed below. This model
describes a cubic Kondo lattice with spinful d-orbital and
f-orbital electrons that are hybridized by inter-orbital coupling.
The basis function for this four band model is:
$|\Psi\rangle=(d_{\ua},d_{\da},f_{\ua},f_{\da})^T$
\cite{legner2014}. The bulk model shows 3D topological crystalline
insulating phases with a non-trivial mirror Chern number at
certain mirror invariant planes \cite{legner2014,legner2015}. Here we consider the thin film
configuration stacking along the direction perpendicular to the
mirror invariant plane. The intralayer part of Hamiltonian ($H_{00}$) and interlayer
part of Hamiltonian ($H_{01}$) are given by \be
H_{00}=\begin{pmatrix}
h^d_{00} & \Phi_{00} \\
\Phi_{00} & h^f_{00}
\end{pmatrix},\
H_{01}=\begin{pmatrix}
h^d_{01} & \Phi_{01} \\
\Phi_{10}^{\dagger} & h^f_{01}
\end{pmatrix}\label{Eq:H00_H01}
\ee
where
\bea
h^{d}_{00}&=&(-2t_{d}(c_1+c_2)-4t_{d}'c_1c_2)\sigma_0 \nonumber \\
h^{f}_{00}&=&(e_f-2t_{f}(c_1+c_2)-4t_{f}'c_1c_2)\sigma_0 \nonumber \\
h^{d(f)}_{01}&=&(-t_{d(f)}-2t_{d(f)}'(c_1+c_2)-4t_{d(f)}''c_1c_2)\sigma_0 \nonumber \\
\Phi_{00}&=&-2[s_1\sigma_x(V_1+V_2c_2)+s_2\sigma_y(V_1+V_2c_1)] \nonumber \\
\Phi_{01}&=&-[s_1V_2\sigma_x+s_2V_2\sigma_y+\frac{1}{i}\sigma_z(V_1+V_2(c_1+c_2))]. \nonumber \\
&&
\eea We label $c_i=\cos k_i$ and $s_i=\sin k_i$ with $i=1,2,3$ for
short. We only focus on the monolayer case (described by $H_m$)
and the bilayer case (described by $H_b$) with the corresponding
Hamiltonians given by \bea
&&H_{m}=H_{00}, \label{Eq:H_m}\\
&&H_{b}=\begin{pmatrix}
H_{00} & H_{01} \\
H_{01}^{\dagger} & H_{00}
\end{pmatrix},
\label{Eq:H_m and H_b} \eea respectively. In general, an $n$-layer
($n=1,2,3,...$) thin film model can be constructed in a similar way.

%\subsection{Mirror symmetry, mirror Chern number and topological phase transition}
Before we study the details of topological phase transition in
monolayer and bilayer films of TMKI, we first discuss symmetry
properties for our system. For a 2D thin film, out-of-plane mirror
symmetry $m_z$ plays a central role in protecting topological
crystalline phases. Under the basis $|\Psi\rangle$, the bulk
mirror operation is $m_z=i\tau_z\otimes\sigma_z$ with $\tau$ and
$\sigma$ for the orbital and spin degree of freedom. Starting from
the bulk mirror operation $m_z$, it is straightforward to write
down the mirror operations for monolayer model ($M_z^m$) and
bilayer model ($M_z^b$) as \bea
&M_z^{m}=&m_z=i\tau_z\otimes\sigma_z \nonumber \\
&M_z^{b}=&\alpha_x\otimes
m_z=i\alpha_x\otimes\tau_z\otimes\sigma_z, \eea where $\alpha_i$
denotes the Pauli matrix of layer degree of freedom in the bilayer
case. In a mirror symmetric system, all the eigen-states can be
characterized by their mirror parities (the eigen-values of mirror
operators) and two subspaces characterized by opposite mirror
parities are decoupled. When time-reversal symmetry is present, a non-zero
net Chern number $C$ is forbidden, while mirror symmetry allows a
possible non-vanishing Chern number $C_{\pm i}$ in each mirror
subspace (with mirror eigenvalue being either $+i$ or $-i$), leading
to the topological mirror insulator phase. In general, we have
\bea
&C&=C_{+i}+C_{-i}=0 \nonumber \\
&C_m&=(C_{+i}-C_{-i})/2=C_{+i}. \eea Therefore, to understand the
topology of the whole system (with time-reversal symmetry), it is
sufficient for us to check only the Chern number in one mirror
subspace (for example, $C_{+i}$ for the $+i$ mirror subspace).

Here we would like to point out an interesting emergent symmetry
that is usually ignored in previous discussions of topological
mirror insulators. With mirror symmetry, since two mirror
subspaces are decoupled, the full Hamiltonian is block diagonal
under the basis with definite mirror parities.
%we can always find a unitary transformation that makes the non-interacting Hamiltonian block-diagonal, with each sub-block corresponding to one mirror subspace. In this case,
As a result, we can perform different $U(1)$ phase transformations
on each mirror subspace, while leaving the Hamiltonian
invariant under such operation. In analogy to spin $U(1)$ symmetry, we
call this emergent $U(1)_m$ symmetry as a ``pseudo-spin" symmetry.
%This means that charge in each mirror subspace (or equivalently mirror parity) will be conserved. This conservation leads to an emergent $U(1)_m$ symmetry and we call it a pseudo-spin symmetry.
The mirror symmetry and pseudo-spin symmetry $U(1)_m$ are
essentially different, but they cannot be distinguished in the
non-interacting limit. Their difference will be clarified when we
consider interaction effects in a later section.
%While later when we consider the interactions in this system, this difference will lead to very important physical consequences.

\subsection{Phase diagram of a monolayer system}

In this section we discuss possible topological phase transitions
(TPTs), as well as the phase diagram, as a function of the
off-block-diagonal coefficients $V_1$ and $V_2$ in Eq.
(\ref{Eq:H00_H01}), which control the hybridization
 gap between d-electrons and f-electrons, in a monolayer TMKI thin film.
%According to Eq. [\ref{Eq:H_m}], since the hybridization gap between
%d electron and f electron is only controlled by the
%off-block-diagonal coefficients $V_1$ and $V_2$ in Eq.
%[\ref{Eq:H00_H01}], possible topological phase
%transitions (TPTs) are also induced by tuning $V_1$ and $V_2$.
%For the monolayer Hamiltonian $H_m$,
We find that TPTs are determined by two conditions  \bea
&&h^d_{00}=h^f_{00} \label{Eq:Condition1} \\
&&\Phi_{00}(k_x,k_y)=0  \label{Eq:Condition2} \eea in $H_{00}$
(Eq. (\ref{Eq:H_m})). We generally search for solutions for the above two
equations at high symmetry lines in the Brillouin zone (BZ). The explicit expression for
Eq. (\ref{Eq:Condition1}) is written as \be
2(t_d-t_f)(c_1+c_2)+4(t_d'-t_f')c_1c_2+e_f=0, \ee
which is simplified to \be 6(c_1+c_2)-6c_1c_2-5=0, \label{Eq:Condition11}
\ee with the parameters
$t_d=1,t_d'=-0.5,e_f=-2,t_f=-\frac{1}{5}t_d,
t_f'=-\frac{1}{5}t_d'$ according to Ref.
[\onlinecite{legner2015}].
%According to Ref. [\onlinecite{legner2015}], the parameters are
%chosen to be $t_d=1,t_d'=-0.5,e_f=-2,t_f=-\frac{1}{5}t_d,
%t_f'=-\frac{1}{5}t_d'$ to reproduce the inverted band structure in
%SmB$_6$.
%(\textcolor{red}{Here it is simply a model study in the original paper, and the unit is arbitrary unit. I think it might be better to say that these parameters reproduce the correct band inversions.})
%(\textcolor{red}{This seems to be a very technical question, and I don't know the answer. We can argue that this set parameter fits the bulk band structure, while other parameters might be unrealistic.})
%Eq. [\ref{Eq:Condition1}] becomes \be
%12(c_1+c_2)-12c_1c_2-2=0. \label{Eq:Condition11} \ee
Eq. (\ref{Eq:Condition11}) can be solved and we find the following
solutions: (1) $(k_x,k_y)=(\pi,\cos^{-1}\frac{11}{12})$ along the
$X$-$M$ line ($k_x=\pi,k_y=k$); (2)
$(k_x,k_y)=(\cos^{-1}(1-\frac{\sqrt{6}}{6}),\cos^{-1}(1-\frac{\sqrt{6}}{6}))$
along the $M$-$\Gamma$ line ($k_x=k_y=k$); (3)
$(k_x,k_y)=(\cos^{-1}\sqrt{\frac{5}{6}},\pi-\cos^{-1}\sqrt{\frac{5}{6}})$
along the $Y$-$X$ line ($k,\pi-k$). Here we only list the
solutions in the first quarter of the BZ and the
solutions in other parts of the BZ can be obtained by performing
four-rotation rotation. On the other hand, Eq. (\ref{Eq:Condition2}) can be
simplified as \bea
&\sin k_x (V_1+V_2 \cos k_y )&=0 \nonumber \\
&\sin k_y (V_1+V_2\cos k_x)&=0, \eea which can be satisfied by
$V_1=V_2$ for the $X$-$M$ line and
$V_1=-(1-\frac{\sqrt{6}}{6})V_2$ at
$(k_x,k_y)=(\cos^{-1}(1-\frac{\sqrt{6}}{6}),\cos^{-1}(1-\frac{\sqrt{6}}{6}))$
along the $M$-$\Gamma$ line.
 %with the help of our results from Eq. (\ref{Eq:Condition11}).
% From this condition, TPT can not occur at $Y$-$X$ line. At $X$-$M$, we require $V_1=V_2$, while at $M$-$\Gamma$, we require $V_1=-(1-\frac{\sqrt{6}}{6})V_2$.
Combining the solutions from the Eq. (\ref{Eq:Condition1}) and
(\ref{Eq:Condition2}) leads to two types of TPTs: (1) When
$V_1=V_2$, TPT happens at $(k_x,k_y)=(\pi,\cos^{-1}\frac{11}{12})$
along the $X$-$M$ line; (2) When $V_1=-(1-\frac{\sqrt{6}}{6})V_2$,
TPT happens at
$(k_x,k_y)=(\cos^{-1}(1-\frac{\sqrt{6}}{6}),\cos^{-1}(1-\frac{\sqrt{6}}{6}))$
along the $\Gamma$-$M$ line. Because of the four-fold rotation
symmetry, each of the TPTs will occur at four different momenta
simultaneously.

%At $\Gamma$-$X$
%line ($k_x=k,k_y=0$), it is easy to show that Eq.
%[\ref{Eq:Condition11}] has no solution. At $X$-$M$ line
%, the solution is $(k_x,k_y)=(\pi,\cos^{-1}\frac{11}{12})$.
%At $M$-$\Gamma$ line ($k_x=k_y=k$), the solution is
%$(\cos^{-1}(1-\frac{\sqrt{6}}{6}),\cos^{-1}(1-\frac{\sqrt{6}}{6}))$.
%At $Y$-$X$ line ($k,\pi-k$), the solution is
%$(\cos^{-1}\sqrt{\frac{5}{6}},\pi-\cos^{-1}\sqrt{\frac{5}{6}})$.
%These results are only for a quarter of the Brillouin zone (BZ),
%while the rest of the BZ has the same behavior because of four-fold rotation symmetry in the system.
\begin{figure}[t]
  \centering
\includegraphics[width=3.5 in]{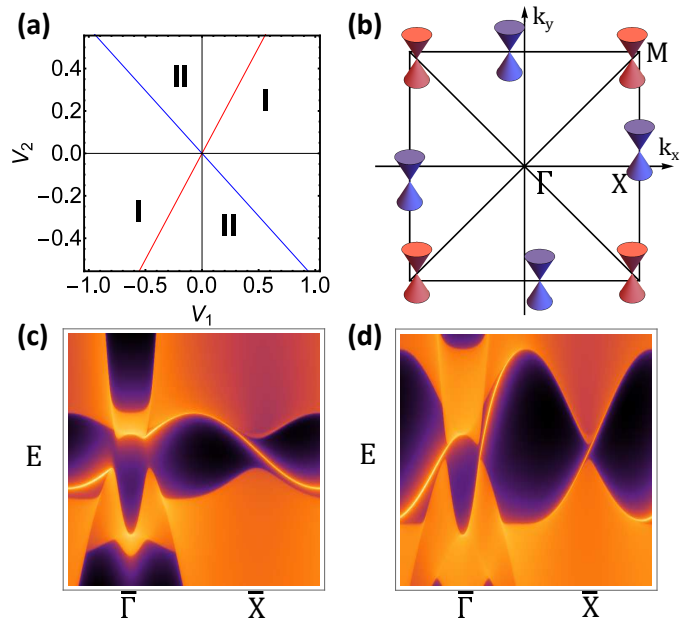}
\caption{Topological phase diagram in the monolayer system. Red
(blue) line represents TPT happening at $X$-$M$ ($\Gamma$-$M$) line.
In Fig. (b), the blue (red) Dirac cone represents the position of
a TPT characterized by the blue (red) critical line in (a). Region
I is a TCI phase with mirror Chern number $-1$, while region II is
a TCI phase with mirror Chern number $+3$. Edge dispersions are
calculated using iterative Green function method for a
semi-infinite configuration: (c) Region I with $C_m=-1$, (d)
Region II with $C_m=+3$. }
  \label{Fig:Phase diagram single layer}
\end{figure}

%In our system, the presence of spatial inversion symmetry and TR symmetry leads to an interesting feature of the TPT.
%Notice that TR symmetry operation $\Theta$, together with inversion
%symmetry operation $P$, guarantees doubly degenerate energy states throughout the BZ. Both symmetry operations commute with mirror symmetry operation $M_z^{m,b}$. It is easy to show that for mirror
%eigenstate $M_z^{m,b}|\pm i,k\rangle=\pm i|\pm i,k\rangle$, one
%has \bea
%&M_z^{m,b}(\Theta|\pm i,k\rangle)&=\mp i(\Theta|\pm i,k\rangle) \nonumber \\
%&M_z^{m,b}(P|\pm i,k\rangle)&=\pm i(\Theta|\pm i,k\rangle) \nonumber \\
%&M_z^{m,b}(P\Theta|\pm i,k\rangle)&=\mp i(P\Theta|\pm i,k\rangle).
%\eea Since the state $P\Theta|\pm i,k\rangle$ carries the momentum
%$+k$ but mirror parity $\mp i$, it is degenerate with $|\pm i,k\rangle$, but with an opposite mirror parity. As a result, any
%TPT occurs simultaneously in both mirror subspaces. We can conclude that the change in mirror Chern
%number is four across the TPTs discussed above.

The phase diagram is summarized in Fig. \ref{Fig:Phase diagram single layer}
(a), where the red line shows the TPTs at the $X$-$M$ line while
the blue line is for the TPTs at the $\Gamma$-$M$ line. The blue
(red) Dirac cones in Fig. \ref{Fig:Phase diagram single layer} (b)
depict the exact positions of the TPTs represented by the blue
(red) line in Fig. \ref{Fig:Phase diagram single layer} (a). To
identify the mirror Chern number in the region I and II, we
calculate the edge state dispersion in a ribbon configuration
along the $y$ direction for the mirror subspace with the mirror
parity $+i$ in Fig. \ref{Fig:Phase diagram single layer} (c) and
(d). The number and chirality of chiral edge states correspond to
both the absolute value and the sign of bulk mirror Chern number
according to the bulk-boundary correspondence. The phase in the
region I (Fig. \ref{Fig:Phase diagram single layer} (c)) possesses
mirror Chern number $C_m=C_{+i}=-1$, while that in the region II
(Fig. \ref{Fig:Phase diagram single layer} (d)) carries
$C_m=C_{+i}=3$. The change of the Chern number $|\Delta C_m|=4$
across the TPTs is due to the four-fold rotation symmetry of the
system.

\subsection{Phase diagram of a bilayer system}

In the monolayer system, TPTs are controlled by hybridization
effects. In contrast, an additional ingredient, the QCE, also
plays a role in the TPT of a bilayer system. QCE is determined by
the inter-layer hopping (the off-block-diagonal term) of the
Hamiltonian. To make connection between monolayer and bilayer
films, we introduce an inter-layer coupling parameter $\lambda\in
[0,1]$ between two layers and re-write the Hamiltonian as \be
H_{b}(\lambda)=\begin{pmatrix}
H_{00} & \lambda H_{01} \\
\lambda H_{01}^{\dagger} & H_{00}
\end{pmatrix}.
\label{Eq:H(V)} \ee When $\lambda=0$, two layers are completely
decoupled, and $H_{b}(0)$ describes the monolayer system with the
phase diagram shown in Fig. \ref{Fig:Phase diagram single layer} (a). When
$\lambda=1$, two layers are strongly coupled, and $H_{b}(1)$
reproduces the bilayer system $H_b$ in Eq. (\ref{Eq:H_m and H_b}).
Therefore, by tuning $\lambda$ continuously, we can understand the
evolution of the energy spectrum from the monolayer system to the
bilayer system.
%Now a naive expectation is that by tuning $V$ continuously, we should be able to understand how QCE itself influences the evolution from monolayer TPD to bilayer TPD. Together with our knowledge of hybridization effect learned from monolayer systems, TPD in bilayer systems seem to have a well-defined and straightforward solution.
%However, QCE and hybridization effect are not independent from each other. This is because the inter-layer hopping term $H_{01}$ also has contribution from the hybridization terms. Therefore, tuning $V$ also changes hybridization effect in some sense. The competition between QCE and hybridization effect actually gives rise to a much richer TPD in a bilayer system than that of a monolayer system. Meanwhile, the complication of solving TPD increases, making it difficult for us to apply analytical approaches.

We emphasize that the $M_z^b$ mirror parity of an eigen-state in
the bilayer film is not directly related to the $M_z^m$ mirror
parity for an eigen-state in each monolayer. A nonzero $\lambda$
introduces interlayer coupling, resulting in the formation of
bonding and anti-binding states. Let us denote $|+i (-i),
n\rangle$ as the mirror even (odd) state of the monolayer mirror
operation $M_z^m$ for the $n$th layer ($n=1,2$). For the bilayer
system, the even and odd states under $M_z^b$ are the linear
combinations of $|\pm i, n\rangle$: \bea
\langle M_z^b\rangle=+i &:& \frac{1}{\sqrt{2}}(|+i,1\rangle+|+i,2\rangle) \nonumber\\
&&\frac{1}{\sqrt{2}}(|-i,1\rangle-|-i,2\rangle ) \nonumber \\
\langle M_z^b\rangle=-i &:& \frac{1}{\sqrt{2}}(|-i,1\rangle+|-i,2\rangle) \nonumber\\
&&\frac{1}{\sqrt{2}}(|+i,1\rangle-|+i,2\rangle ). \label{Eq:Mirror
eigenstate}\eea Therefore, starting from any $M_z^m$ mirror
eigen-state, one can obtain both even and odd $M_z^b$ mirror eigen
state of the bilayer system by making linear combination of the
states in two layers.

\begin{figure}[t]
  \centering
\includegraphics[width=2.8 in]{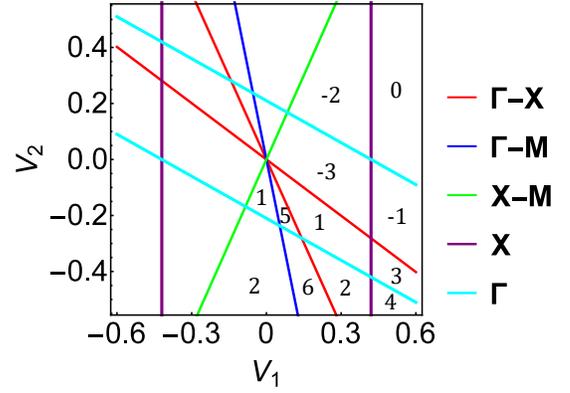}
\caption{Topological phase diagram for the bilayer system. TPT
critical lines are plotted using different color, as shown in
the legend. The numbers shown in the figure are the corresponding
mirror Chern numbers.}
  \label{Fig:Phase diagram double layer}
\end{figure}

We now map out the phase diagram of the bilayer system numerically and mirror
Chern numbers are marked for different phases in Fig.
\ref{Fig:Phase diagram double layer}. As a result of the
combination of QCE and hybridization effect, TPTs can occur either
along some high symmetry lines (e.g. $\Gamma$-$X$, $\Gamma$-$M$
and $X$-$M$ lines) or at some high symmetry points (e.g. $X$ and
$\Gamma$), as shown by lines with different colors in Fig.
\ref{Fig:Phase diagram double layer}. Thanks to the four-fold
rotation symmetry of the cubic lattice, TPTs along the lines
$\Gamma$-$X$, $\Gamma$-$M$ and $X$-$M$, labeled by red, blue and
green lines in Fig. \ref{Fig:Phase diagram double layer}, will
change the mirror Chern number by $\pm 4$. In contrast, TPTs at
the high symmetry points $\Gamma$ ($X$) will change the mirror
Chern number by $\pm 1$ ($\pm 2$), which are labeled by the cyan
(purple) lines in Fig. \ref{Fig:Phase diagram double layer}. Due
to the multiple TPTs in the bilayer system, we find that the
mirror Chern number can be as large as $C_m=6$ in the phase diagram that has been mapped out.

%As we have stated, the complexity of TPD is a direct result of competition between QCE and hybridization effect. In the legend, high symmetry points (lines) where a TPT happens is labelled with different colors.

%For the double layer case, it is not easy to analytically solve for the TPT conditions. While numerically, it is always possible to map out the phase diagram, as is shown in Fig. \ref{Fig:Phase diagram double layer}. This TPT phase diagram is very complicated due to the mechanism competition between hybridization effect and quantum confinement effect. Generally speaking, we expect the quantum confinement effect gives rise to TPT that exactly happens at the high symmetry points. For the TPT due to hybridization effect, based on our experience for the single-layer case, we expect the TPT happens along the high symmetry lines but off the high symmetry points. This conjecture is verified in the TPT phase diagram in Fig. \ref{Fig:Phase diagram double layer}, where both mechanism competes. As we have argued, those TPT off high symmetry points will change the mirror Chern number by $\pm 4$. The TPT happening at $\Gamma$ ($X$) will change the mirror Chern number by $\pm 1$ ($\pm 2$). What is interesting is that, TCI phases with {\bf VERY LARGE} mirror Chern numbers can be achieved by simply tuning $V_{1,2}$ for the double layer system.

\begin{figure}[t]
  \centering
\includegraphics[width=3.5 in]{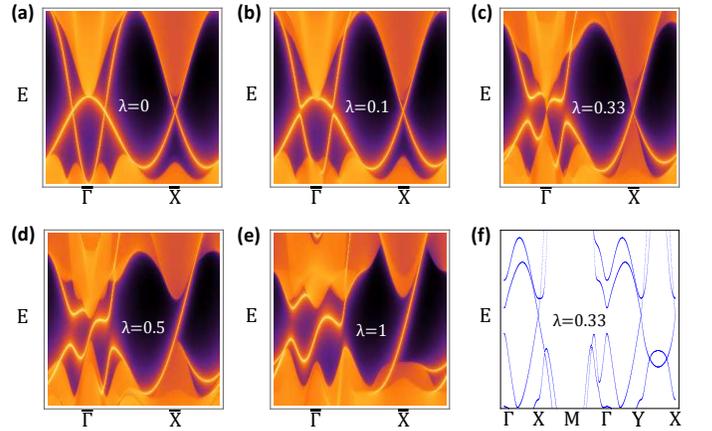}
\caption{Edge dispersion plots of the bilayer system with a different layer coupling $\lambda$
are plotted in (a) to (e). In (f), we plot the bulk dispersion
exactly at the topological phase transition.}
  \label{Fig:layer transition}
\end{figure}

As an example, we track the formation of a $C_m=2$ TCI phase (e.g.
$V_1=0,V_2=-0.3$) based on our layer construction scheme by
varying $\lambda$ from $0$ to $1$ in Eq. (\ref{Eq:H(V)}). Here we
choose the same set of parameters for each monolayer that belongs
to the region II in Fig. \ref{Fig:Phase diagram single layer} with
a mirror Chern number $C_m=+3$. As discussed above, the eigen
state with a definite mirror parity in the bilayer system is a
linear combination of the eigen-states with both even and odd
mirror parities in the monolayer systems. As a result, for a small
$\lambda$ ($\lambda=0.1$), we find a trivial phase
($C_m=0$) in the new $+i$ mirror subspace as shown in Fig.
\ref{Fig:layer transition} l(b) (The gapless edge modes in
Fig. \ref{Fig:layer transition} (a) are because two layers are
decoupled for $\lambda=0$ in the bilayer system). By tuning
$\lambda$, the system undergoes a TPT at $\lambda=0.33$, where the
bulk band gap closes at both $X$ and $Y$ in the BZ, as shown in
Fig. \ref{Fig:layer transition} (f). As a consequence, the mirror
Chern number $C_m$ is changed by 2, giving rise to the topological
mirror insulator phase with $C_m=2$ for $\lambda>0.33$.
%The corresponding edge dispersion and bulk
%dispersion of $+i$ mirror subspace are shown in Fig.
%\ref{Fig:layer transition} (c) and (f). In (c), bulk gap closes at
%$\Gamma$ because $Y$ point is projected onto the $\Gamma$ point in
%the effective edge BZ. This is the only TPT during the evolution. It changes $C_m$ by 2, and gives
%rise to the topological mirror insulator phase with $C_m=2$ at $\lambda=1$.
As shown in Fig. \ref{Fig:layer transition} (d) and (e) for
$\lambda=0.5$ and $\lambda=1$, two chiral gapless edge modes exist
in the mirror parity subspace $+i$ with one of them around
$\bar{\Gamma}$ (projected from $Y$ in the bulk BZ), while the
other around $\bar{X}$ in the edge BZ. Taking into account the
other mirror parity subspace ($-i$), two copies of helical edge
modes exist in this bilayer system for $\lambda=1$.
%({\bf I'm not sure if people call the edge modes of topological mirror insulators as helical modes or other names, could you please check it? }) (\textcolor{red}{Other references of 2D TCIs usually call this spin-filtered edge states, for example, in Junwei Liu and Liang Fu's paper. I think we can call them helical, since spin and momentum are locked? })

To conclude, we have demonstrated the bilayer model of a TMKI as a
playground for 2D topological mirror insulator phases with various
mirror Chern numbers. Different from either monolayer model or
bulk model, the richness of topological mirror insulator phases in bilayer TMKI
originates from the interplay between hybridization effect and
QCE. Our layer construction scheme can be generalized to
multiple-layer systems and thus, topological phases with higher
mirror Chern numbers are expected to exist and can be tuned by the
thickness of thin films. Next we will study the interaction effect
in the bilayer system with the mirror Chern number $C_m=2$ (two
copies of helical edge modes).

\section{Interacting edge states of 2D TMKIs}

In this section, we will discuss interacting physics at the 1D
edge of 2D TMKIs. With the presence of interactions, previous
studies have shown that the topological classification of 2D TMKI will
be reduced from $\mathbb{Z}$ to $\mathbb{Z}_4$
\cite{isobe2015,yoshida2015} in the case that the pseudo-spin
symmetry $U(1)_m$ is broken while the mirror symmetry is preserved
by the interaction.
%This means that
%mirror Chern number is only well defined modulo $4$.
Similar change of topological classification should occur in our
phase diagram (Fig. (\ref{Fig:Phase diagram double layer})). Our
proposed TMKI thin film system with high mirror Chern numbers
provides us an ideal platform to test the reduction of topological
classification both theoretically and experimentally.

In this section, our main interest focuses on another intriguing
aspect of interacting topological phases. In a recent
paper~\cite{bi2016}, it was proposed that a BSPT phase protected
by $U(1)_c\times U(1)_s$ symmetry can be realized in a bilayer graphene
system under Coulomb interaction and a strong magnetic field. Here
$U(1)_c$ and $U(1)_s$ denote $U(1)$ symmetries corresponding to
charge conservation and spin conservation, respectively. Graphene
under a strong magnetic field becomes a quantum spin Hall
insulator~\cite{abaninlee}, which was recently demonstrated
experimentally~\cite{young2013}. Due to the similarities between a
quantum spin Hall insulator and a 2D topological mirror insulator,
it is natural to ask whether BSPT states could occur in our
systems. With Abelian bosonization, we will discuss possible
realization of BSPT states, as well as other related interacting
phases. For this purpose, we will focus on a 2D TMKI phase with
the mirror Chern number $C_m=2$ in the rest of our discussions.

\subsection{Abelian bosonization of interacting edge states of 2D TMKIs}

The low energy physics at the edge of a $C_m=2$ TMKI is well
captured by the following non-interacting two-channel helical
Luttinger liquid model \be
H_0=\frac{v_f}{2}\sum_{l=1,2}[\psi^{\dagger}_{l,L}i\partial_x\psi_{l,L}-\psi^{\dagger}_{l,R}i\partial_x\psi_{l,R}],
\label{Eq:Luttinger model} \ee with the channel index $l=1,2$. As shown in Fig. \ref{Fig:layer transition}
(e), we can define $l$ as the valley index and label the
helical edge modes at $\bar{\Gamma}$
($\bar{X}$) with $l=1$ ($l=2$). $L$
($R$) denotes the left mover (right mover) of chiral fermions in
the even (odd) mirror subspace since mirror parity is locked to
the direction of velocity of the edge states. The Abelian
bosonization of the Hamiltonian (Eq. (\ref{Eq:Luttinger model}))
has been discussed in Ref.~[\onlinecite{bi2016}]. To keep the
current paper self-contained, we briefly review the bosonization
scheme as follows: \bea
\psi_{l,R} &\sim& e^{i2\sqrt{\pi}\chi_{l,R}} \nonumber \\
\psi_{l,L} &\sim& e^{-i2\sqrt{\pi}\chi_{l,L}},
\label{Eq:bosonization scheme_I} \eea where $\chi$ are chiral
bosonic fields. We introduce bosonic dual variables $\theta_l$ and
$\phi_l$ as \bea
\phi_l&=&\chi_{l,R}+\chi_{l,L} \nonumber \\
\theta_l&=&-\chi_{l,R}+\chi_{l,L}, \label{Eq:dual fields} \eea and
write the fermionic density operators in terms of dual variables
as \bea
\rho_{R,l}&=&\frac{1}{2\sqrt{\pi}}\partial_x(\phi_l-\theta_l) \nonumber \\
\rho_{L,l}&=&\frac{1}{2\sqrt{\pi}}\partial_x(\phi_l+\theta_l).
\label{Eq:density operators} \eea Finally, we define the bonding
and anti-bonding states between different channels as \bea
\phi_+&=&\frac{1}{\sqrt{2}}(\phi_1+\phi_2),\ \phi_-=\frac{1}{\sqrt{2}}(\phi_1-\phi_2) \nonumber \\
\theta_+&=&\frac{1}{\sqrt{2}}(\theta_1+\theta_2),\ \theta_-=\frac{1}{\sqrt{2}}(\theta_1-\theta_2).
\label{Eq:bonding and anti-bonding}
\eea

%A general Coulomb interaction is given below:
%\bea
%H_{int}&=&H_1+H_2+H_\text{exchange} \nonumber \\
%&=&\sum_{i=1,2}U_0\rho_{i,\ua}\rho_{i,\da}+\sum_{i\neq j,\sigma,\sigma'}U_1\rho_{i,\sigma}\rho_{j,\sigma'} \nonumber \\
%&=&\sum_{i=1,2}U_0\rho_{i,L}\rho_{i,R}+\sum_{i\neq j}U_1(\rho_{i,R}\rho_{j,L}+\rho_{i,L}\rho_{j,R})
%\eea
%Here we have ignored the forward scattering terms ($\sigma=\sigma'$) in $H_2$. Translate $H_{int}$ in the bosonization language, we have
The free fermion Hamiltonian (Eq. (\ref{Eq:Luttinger model})) can
be transformed into a free boson Hamiltonian in terms of bosonic
variables $\phi_\pm$ and $\theta_\pm$. In the bosonic Hamiltonian,
two-body interaction terms, which can be explicitly written in
terms of density operators, can renormalize Fermi velocities and
Luttinger parameters. These interaction terms include \bea
H_1&=&\sum_{l=1,2}g_1\rho_{l,L}\rho_{l,R} \nonumber \\
%&=&\frac{g_1}{4\pi}\sum_{l=1,2}\partial_x(\phi_l-\theta_l)\partial_x(\phi_l+\theta_l) \nonumber \\
%&=&\frac{g_1}{4\pi}\sum_{l=1,2}[(\partial_x\phi_l)^2-(\partial_x\theta_l)^2] \nonumber \\
&=&\frac{g_1}{4\pi}[(\partial_x\phi_-)^2+(\partial_x\phi_+)^2-(\partial_x\theta_-)^2-(\partial_x\theta_+)^2] \nonumber \\
H_2&=&\sum_{i\neq j}g_2(\rho_{i,R}\rho_{j,L}+\rho_{i,L}\rho_{j,R}) \nonumber \\
%&=&g_2\rho_{1,R}\rho_{2,L}+\rho_{1,L}\rho_{2,R} \nonumber \\
%&=&\frac{g_2}{4\pi}[\partial_x(\phi_l-\theta_l)\partial_x(\phi_2+\theta_2)+\partial_x(\phi_2-\theta_2)\partial_x(\phi_1+\theta_l)] \nonumber \\
%&=&\frac{g_2}{2\pi}[\partial_x\phi_1\partial_x\phi_2-\partial_x\theta_1\partial_x\theta_2] \nonumber \\
&=&\frac{g_2}{4\pi}[(\partial_x\phi_+)^2-(\partial_x\phi_-)^2+(\partial_x\theta_-)^2-(\partial_x\theta_+)^2]. \nonumber \\
&& \label{Eq:H_1 and H_2} \eea Together with Eq.
(\ref{Eq:Luttinger model}), the full harmonic Hamiltonian of
bosons is given by \bea
H_\pm&=&\frac{v_\pm}{2}[K_\pm(\partial_x\phi_\pm)^2+\frac{1}{K_\pm}(\partial_x\theta_\pm)^2]
%H_-&=&\frac{v_-}{2}[K_-(\partial_x\phi_-)^2+\frac{1}{K_-}(\partial_x\theta_-)^2] \nonumber \\
%H_+&=&\frac{v_+}{2}[K_+(\partial_x\phi_+)^2+\frac{1}{K_+}(\partial_x\theta_+)^2],
\label{Eq:renormalized H_0}
\eea
with the corresponding Luttinger parameters
\bea
K_-&=&\sqrt{\frac{v_f+\frac{g_1}{2\pi}-\frac{g_2}{2\pi}}{v_f-\frac{g_1}{2\pi}+\frac{g_2}{2\pi}}} \nonumber \\
K_+&=&\sqrt{\frac{v_f+\frac{g_1}{2\pi}+\frac{g_2}{2\pi}}{v_f-\frac{g_1}{2\pi}-\frac{g_2}{2\pi}}}.
\label{Eq:Luttinger parameter}
\eea

Now let us consider scattering that corresponds to anharmonic
terms in the Hamiltonian. These terms include \bea
H_{\alpha_1}&=&\alpha_1 \psi^{\dagger}_{1,L}\psi_{1,R}\psi^{\dagger}_{2,R}\psi_{2,L}+h.c. \nonumber \\
&\sim& \alpha_1 \cos{2\sqrt{2\pi}\phi_-} \nonumber \\
H_{\alpha_2}&=&\alpha_2 \psi^{\dagger}_{1,L}\psi_{2,R}\psi^{\dagger}_{1,R}\psi_{2,L}+h.c. \nonumber \\
&\sim& \alpha_2 \cos{2\sqrt{2\pi}\theta_-}, \nonumber \\
H_{\alpha_3}&=&\alpha_3 \psi^{\dagger}_{1,L}\psi_{1,R}\psi^{\dagger}_{2,L}\psi_{2,R}+h.c. \nonumber \\
&\sim& \alpha_3\cos{}2\sqrt{2\pi}\phi_+.
%H_{\beta_2}&=&\beta_2 \psi^{\dagger}_{1,L}\psi_{1,R}\psi^{\dagger}_{1,L}\psi_{1,R}+\psi^{\dagger}_{2,L}\psi_{2,R}\psi^{\dagger}_{2,L}\psi_{2,R}+h.c. \nonumber \\
%&\sim& \beta_2(\cos{4\sqrt{\pi}\phi_1}+\cos{4\sqrt{\pi}\phi_2}). \nonumber \\
\label{Eq:Umklapp scattering} \eea
%It should be pointed out that
%$H_{\alpha_2}$ and $H_{\alpha_3}$ describe Umklapp scattering
%process and break translation symmetry ({\bf We should define
%translation symmetry explicitly here, or maybe define U(1)$_m$
%here explicitly}).
In summary, the full form of Hamiltonian is
given by \bea
H&=&\sum_{l=\pm}\frac{v_l}{2}[K_l(\partial_x\phi_l)^2+\frac{1}{K_l}(\partial_x\theta_l)^2]+\alpha_1\cos{2\sqrt{2\pi}\phi_-} \nonumber \\
&&+\alpha_2 \cos{2\sqrt{2\pi}\theta_-}+\alpha_3\cos{2\sqrt{2\pi}\phi_+}\nonumber \\
%&&+\beta_2(\cos{4\sqrt{\pi}\phi_1}+\cos{4\sqrt{\pi}\phi_2}).
\label{Eq:Total Ham}
\eea

\subsection{Conditions for BSPT phase}

We first consider symmetry operations of our Hamiltonian in more
details. The charge conservation $U(1)_c$, pseudo-spin symmetry
$U(1)_m$, time-reversal symmetry $\cal{T}$ and out-of-plane mirror
symmetry $M_z$ can be defined as \bea &U(1)_c(\alpha)
\begin{pmatrix}
\phi_l \\
\theta_l
\end{pmatrix}&=
\begin{pmatrix}
\phi_l \\
\theta_l-\frac{\alpha}{\sqrt{\pi}}
\end{pmatrix} \nonumber \\
&U(1)_m(\alpha)
\begin{pmatrix}
\phi_l \\
\theta_l
\end{pmatrix}&=
\begin{pmatrix}
\phi_l-\frac{\alpha}{\sqrt{\pi}} \\
\theta_l
\end{pmatrix} \nonumber \\
&\cal{T}\begin{pmatrix}
\phi_l \\
\theta_l
\end{pmatrix}&=
\begin{pmatrix}
-\phi_l+\frac{\sqrt{\pi}}{2} \\
\theta_l-\frac{\sqrt{\pi}}{2}
\end{pmatrix} \nonumber \\
&M_z\begin{pmatrix}
\phi_l \\
\theta_l
\end{pmatrix}&=
\begin{pmatrix}
\phi_l-\frac{\sqrt{\pi}}{2} \\
\theta_l
\end{pmatrix},
\label{Eq:Symmetry} \eea respectively, where $l=1,2$. From Eq.
(\ref{Eq:Symmetry}), it is easy to see that
$\theta_+=\frac{1}{\sqrt{2}}(\theta_1+\theta_2)$ carries the
$U(1)_c$ charge, and $\phi_+$ carries the $U(1)_m$ charge. We find
that $H_{\alpha_1}$ and $H_{\alpha_2}$ preserve all the four
symmetries, while $H_{\alpha_3}$ breaks $U(1)_m$ symmetry and
preserves the mirror symmetry $M_z$. In the Appendix, we will show
that $H_{\alpha_{3}}$ can originate from the standard Coulomb
exchange interaction. Therefore, we conclude that a general
Coulomb interaction reduces the symmetry of the 2D TMKI system from
$U(1)_c\times U(1)_m\times M_z\rtimes \cal{T}$ to $U(1)_c\times
M_z\rtimes \cal{T}$.

The existence of $H_{\alpha_3}$ is the key difference between our
model and the interacting bilayer graphene model in
Ref.~[\onlinecite{bi2016}]. In the bilayer graphene system, the
BSPT phase is protected by $U(1)_c\times U(1)_s$ symmetry. Spin
conservation symmetry $U(1)_s$ is preserved because (1) the spin-orbit
coupling (SOC) effect is negligible and (2) the Coulomb interaction
conserves total spin. In our TMKI system, the pseudo-spin symmetry
$U(1)_m$ is playing the same role as the $U(1)_s$ symmetry in the
bilayer graphene system. However, strong SOC exists in our system
and the Coulomb interaction does not respect the pseudo-spin
symmetry $U(1)_m$ due to the $H_{\alpha_3}$ term (Refer to the
appendix for more details). %As a result, it is essential for us to
%understand the renormalization behavior of $H_{\alpha_3}$ term.
%Naively, one might expect that if the system is tuned away from half-filling, translational symmetry will prohibit both $H_{\alpha_2}$ and $H_{\alpha_3}$. However, a translational breaking term (i.e. disorders) will assist the above Umklapp scattering process to happen.
%Renormalization group (RG) analysis offers us a good approach to understand the effects of $H_{\alpha_3}$.
If $H_{\alpha_3}$ is irrelevant, $U(1)_m$ symmetry will be
recovered under renormalization group (RG) flow. As a result, the
$U(1)_c\times U(1)_m$ symmetry protected BSPT phase will emerge,
in analogy to the case studied in Ref. [\onlinecite{bi2016}]. If
$H_{\alpha_3}$ is relevant, the $U(1)_c\times U(1)_m$ symmetry
will be explicitly broken even at low energy. However, one might
still wonder whether the remaining mirror symmetry $M_z$ can play
the same role as the $U(1)_m$ in protecting a BSPT phase. Next we
will study the RG flow of the $H_{\alpha_{1,2,3}}$ terms.

%To start with, let us briefly review the theory of the 1D
%Sine-Gordan Hamiltonian \be
%H_{SG}=\frac{v_l}{2}[K_l(\partial_x\phi_l)^2+\frac{1}{K_l}(\partial_t\phi_l)^2]+u\cos(m_l\phi_l),
%\ee with the scaling dimension of the cosine term \be
%\Delta(u)=\frac{m_l^2}{4\pi K_l}. \ee and the tree-level beta
%function ($D=2$ in our case) \be \beta(u)=[D-\Delta(u)]u+... \ee
%The $u$ term will be relevant (irrelevant) if $\Delta(u)<2$
%($\Delta(u)>2$).

In our TMKI system, the decoupled Hamiltonians for $\phi_-$ and
$\phi_+$ are \bea
H_{\phi_-}&=&\frac{v_-}{2}[K_-(\partial_x\phi_-)^2+\frac{1}{K_-}(\partial_t\phi_-)^2]+\alpha_1\cos(2\sqrt{2\pi}\phi_-), \nonumber \\
H_{\phi_+}&=&\frac{v_+}{2}[K_+(\partial_x\phi_+)^2+\frac{1}{K_+}(\partial_t\phi_+)^2]+\alpha_3\cos(2\sqrt{2\pi}\phi_+). \nonumber \\
&&
\eea
The scaling dimensions of $\alpha_1$ and $\alpha_3$ are
\bea
\Delta(\alpha_1)&=&\frac{(2\sqrt{2\pi})^2}{4\pi K_-}=\frac{2}{K_-}, \nonumber \\
\Delta(\alpha_3)&=&\frac{(2\sqrt{2\pi})^2}{4\pi K_+}=\frac{2}{K_+}.
\label{Eq:scaling dim_alpha} \eea
From Eq. (\ref{Eq:Luttinger
parameter}), we find that when $g_1>g_2>0$, both $K_-$ and $K_+$
are greater than $1$. Therefore, both interaction terms ($H_{\phi_\pm}$) are
relevant. In the strong coupling limit, $\phi_-$ and $\phi_+$
are pinned to
\bea
\phi_+&=&\frac{(2n+1)\pi}{2\sqrt{2\pi}} \nonumber \\
\phi_-&=&\frac{(2m+1)\pi}{2\sqrt{2\pi}},
\eea
%certain values so that the corresponding cosine
%terms are minimized. The pinning values of $\phi_-$ and $\phi_+$ are
%\bea
%\frac{1}{\sqrt{2}}(\phi_1+\phi_2)&=&\frac{(2n+1)\pi}{2\sqrt{2\pi}} \nonumber \\
%\frac{1}{\sqrt{2}}(\phi_1-\phi_2)&=&\frac{(2m+1)\pi}{2\sqrt{2\pi}},
%\eea
with $n,m\in\mathbb{Z}$ to minimize the cosine terms.
Correspondingly, we have \bea
\phi_1&=&\frac{(n+m+1)\sqrt{\pi}}{2} \nonumber \\
\phi_2&=&\frac{(n-m)\sqrt{\pi}}{2}. \eea
By choosing $n=m=0$, we arrive
at $\phi_1=\frac{\sqrt{\pi}}{2},\phi_2=0$. The mirror symmetry
operation will change $\phi_1$ and $\phi_2$ to
$\phi_1'=0,\phi_2'=-\frac{\sqrt{\pi}}{2}$ with $n=-1,m=0$.
Therefore, $(\phi_1,\phi_2)$ and $(\phi_1',\phi_2')$ are two
degenerate states that are connected by mirror symmetry. In this
case, both fermionic and bosonic degrees of freedom are explicitly
gapped on the boundary. We can define a set of order parameter:
\bea
\Delta_{I,1}&=&\langle \psi^{\dagger}_{1,R}\psi_{1,L}\rangle \sim \langle e^{-2i\sqrt{\pi}\phi_1} \rangle \nonumber \\%= e^{-i\sqrt{2\pi}(\phi_+-(-1)^l\phi_-)}
\Delta_{I,2}&=&\langle \psi^{\dagger}_{2,R}\psi_{2,L}\rangle \sim
\langle e^{-2i\sqrt{\pi}\phi_2} \rangle. \eea When $\phi_+$ and
$\phi_-$ fields are pinned, $\Delta_{I,l=1,2}$ acquires a
non-vanishing expectation value which corresponds to the
spontaneous mirror symmetry breaking (MSB).

\begin{figure}[t]
  \centering
\includegraphics[width=3.0 in]{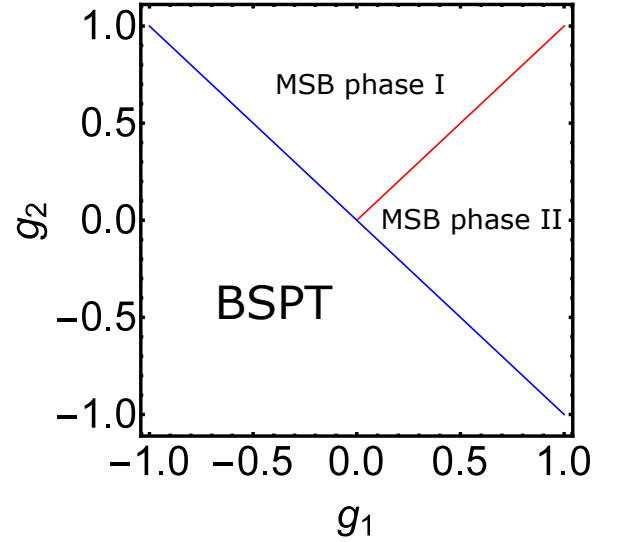}
\caption{Topological phase diagram of BSPT is plotted. Blue line
($K_+=1$) is the topological phase transition that separates the
BSPT phase and trivial phases. Red line is a first order phase
transition that separates two trivial MSB (mirror symmetry
breaking) phases.} \label{Fig:BSPT}
\end{figure}

When $g_2 > g_1 > 0$, $K_- < 1$ and $K_+ > 1$, the terms
$\alpha_3\cos{2\sqrt{2\pi}\phi_+}$ and
$\alpha_2\cos{2\sqrt{2\pi}\theta_-}$ are relevant, which pins
$\phi_+$ and $\theta_-$ to the following values \bea
\phi_+&=&\frac{(2n'+1)\pi}{2\sqrt{2\pi}} \nonumber \\
\theta_-&=&\frac{(2m'+1)\pi}{2\sqrt{2\pi}}, \eea with
$n',m'\in\mathbb{Z}$. This also gaps out all the bosonic degrees
of freedom at the boundary. In this case,
%the above order
%parameter $\Delta_{I,l=1,2}$ cannot correctly characterize the
%symmetry breaking of this system. We could
we need to define a new set of
order parameter as \bea
\Delta_{II,1}&=&\langle \psi^{\dagger}_{1,R}\psi_{2,L}\rangle \sim \langle e^{-i\sqrt{2\pi}(\phi_+-\theta_-)}\rangle \nonumber \\
\Delta_{II,2}&=&\langle \psi^{\dagger}_{2,R}\psi_{1,L}\rangle \sim
\langle e^{-i\sqrt{2\pi}(\phi_++\theta_-)}\rangle, \eea
to characterize this MSB phase.
% In this
%case, this new set of order parameters will obtain nonzero
%expectation values, which also breaks the mirror symmetry
%explicitly.
 %Only $\phi_+$ and $\theta_+$ field will be gapped out, while the other bosonic degree of freedom $\phi_-$ and $\theta_-$ remain gapless. In this case, only $\phi_+$ field will be pinned to certain values. The gapless nature of the anti-binding bosonic fields is not protected by any of those unbroken symmetries. It completely originates from the analysis of RG flow. ({\bf I thought if one includes $\cos\theta_a$ term, $\phi_a$ and $\theta_a$ field will also be gapped %out. })

In general, due to the coexistence of $
\alpha_1\cos{2\sqrt{2\pi}\phi_-} $ and $
\alpha_2\cos{2\sqrt{2\pi}\theta_-} $, the anti-bonding bosonic
degree of freedom is always gapped out for any value of $K_-$
(except for the free case with $K_-=1$). To guarantee the bonding
bosonic mode is gapless, the system should satisfy the condition
\be g_1 + g_2 < 0, \ee which corresponds to attractive
interaction. This condition implies that $H_{\alpha3}$ is
irrelevant. Under this condition, $U(1)_m$ symmetry is recovered
in the infrared limit and can protect a BSPT phase together with the
$U(1)_c$ symmetry, in analogy to the discussion in Ref.
[\onlinecite{bi2016}].

%This concludes that in our 2D TMKI system, time-reversal and
%mirror symmetry alone are not able to support a BSPT phase. ({\bf
%I am not sure I follow this statement. What exactly does it mean
%the TR and mirror together do not support BSPT?})

%From a different perspective, a $U(1)$-like continuous
%symmetry will not be spontaneously broken in 1D, thanks to the
%Mermin-Wagner theorem, while a discrete symmetry (i.e. mirror
%symmetry) is fragile under interactions and can be spontaneously broken.

Based on the discussion above, we have mapped out the phase
diagram of edge states in a TMKI, as shown in Fig.~\ref{Fig:BSPT},
where $K_+=1$ (blue line) separates the gapless edge states and
MSB phases, and thus corresponds to a Kosterlitz-Thouless
transition. In the entire MSB regime, all bosonic degrees of
freedom are gapped out. In addition, this MSB regime can be
further divided into two phases with different
mirror-symmetry-breaking order parameters. The transition line
($K_-=1$) that separates MSB phase I and MSB phase II is a first
order phase transition.

\subsection{Mirror parity domain wall in a mirror symmetry breaking phase}

In the MSB regime, edge states are gapped out as a result of
spontaneous MSB.
%This trivial regime seems to be quite "boring" at the first sight, but it actually hosts exotic physics.
As discussed in Ref. [\onlinecite{chen2014,lu2014}], domain wall
structure of a gapped SPT phase possesses nontrivial degrees of
freedom. In our model, the mirror symmetry as a discrete $Z_2$
symmetry offers us an opportunity to construct a domain wall
between phases which break the mirror symmetry differently. The
mirror symmetry only affect $\phi_+$ field (See Eq.
(\ref{Eq:Symmetry})), \bea M_z\phi_+=\phi_+-\sqrt{\frac{\pi}{2}}.
\label{Eq:Mirror operation} \eea According to the construction
scheme in Ref. [\onlinecite{chen2014}], the domain wall creation
operator $D_M$ is an exponential operator of $\theta_+$ (the dual
field of $\phi_+$): $D_M=e^{iC\theta_+}$.
%To show that $D_M$ is a
%domain wall creation operator and identify the value of $C$, we
%first create a domain wall at the position $x$. We then create a
%field $\phi_+$ at the position $y$ and remove the previous domain
%wall at $x$. Suppose $\phi_+$ is in a MSB state, if $y<x$,
%$\phi_+$ remains at its original MSB state. If $y>x$, $\phi_+$
%will move to a different while degenerate MSB state as a result of
%the domain wall structure.
The effect of the domain wall creation operator is \bea
&&D_M^{-1}(x)\phi_+(y) D_M(x) \nonumber \\
&=&e^{-iC\theta_+}\phi_+e^{iC\theta_+} \nonumber \\
&=&e^{-iC\theta_+}\phi_+[1+\sum_{n=1}^{\infty}\frac{i^nC^n}{n!}(\theta_+)^n] \nonumber \\
&=&\phi_+(y)-C\Theta(x-y). \eea Here we have used the commutation
relation $[\phi_+(x),\theta_+(x')]=i\Theta(x'-x)$ with
$\Theta(x-y)$ being the Heaviside step function. Since the domain
wall operator connects two degenerate MSB vacua, we immediately
obtain $C=\sqrt{\frac{\pi}{2}}$ while comparing with Eq.
(\ref{Eq:Mirror operation}). The complete form of $D_M$ is given
by \be D_M=e^{i\sqrt{\frac{\pi}{2}}\theta_+}. \ee Now we are ready
to explore the properties of the above domain wall operator. The
density operator $j_0$ is given by \be
j_0=\frac{1}{\sqrt{\pi}}\partial_x(\phi_1+\phi_2)=\sqrt{\frac{2}{\pi}}\partial_x\phi_+.
\ee We consider a domain wall at $x_0$ and integrate $j_0$ across
$x_0$ to obtain its charge accumulation of $D_M$ as \be
Q(D_M)=\int_{x_0^-}^{x_0^+}j_0
dx=\int_{x_0^-}^{x_0^+}\sqrt{\frac{2}{\pi}}\partial_x\phi_+ dx=1.
\ee Therefore, $D_M$ carries one unit charge. This can also be
verified by performing the charge $U(1)$ transformation to $D_M$.
On the other hand, we could test its response to TR symmetry
operation $\cal{T}$: \be {\cal T}^2D_M={\cal
T}e^{i\sqrt{\frac{\pi}{2}}\theta_+ +i\frac{\pi}{2}}=-D_M. \ee
Therefore, $D_M$ transforms exactly like a spinful fermion under
${\cal T}$.
%, and carries a half-integer spin.
Thus, we conclude that a domain wall $D_M$ carries a charge-1 spinful
fermion.
%one charge and spin-$\frac{1}{2}$.
%({\bf We need to be careful here: spin is not
% conserved in the system, we cannot really define spin-1/2.})

%However, it is not a real fermion due to its highly non-local nature. To see this ,we express $D_M$ in terms of local fermionic operators:
%\bea
%D_M=(\psi_{1,R}^{\dagger}\psi_{2,L}^{\dagger}\psi_{2,R}^{\dagger}\psi_{2,L}^{\dagger})^{\frac{1}{4}}
%\eea
%Therefore, it is not surprising that $D_M$ has exotic statistics. (\textcolor{red}{To be continued, once I have more thoughts on it...})

\subsection{K matrix formulation of BSPT in interacting TMKIs}
K matrix formulation has successfully been applied to the
classification of SPT phases \cite{lu2012}. In this section, we
connect our discussion based on standard Luttinger liquid
language to the well known K matrix formulation, and explicitly
construct the K matrix for a bosonic SPT phase in our system. To start
with, we introduce a different yet equivalent Abelian bosonization
scheme (compared to Eq. (\ref{Eq:bosonization scheme_I})) as \bea
\psi_{l,R}&\sim& e^{i\chi_{l,R}} \nonumber \\
\psi_{l,L}&\sim& e^{-i\chi_{l,L}}, \eea while the definition of
dual variables are modified as \bea
\phi_{l}&=&\chi_{l,R}+\chi_{l,L} \nonumber \\
\theta_{l}&=&\chi_{l,R}-\chi_{l,L}, \eea with an
extra sign factor in the definition of $\theta_l$ field for future
convenience. In the Lagrangian form, our Hamiltonian is transformed to: \bea {\cal L}=\frac{1}{4\pi}(K_{IJ}\partial_t
\chi_I \partial_x\chi_J - V_{IJ}\partial_x \chi_I
\partial_x\chi_J), \eea where
$\chi=(\chi_{1,R},\chi_{1,L},\chi_{2,R},\chi_{2,L})^T$. Matrix $V$
is determined by the Hamiltonian of the system, the form of which
is not interesting to us. The K matrix of the system is given by
\bea K=
\begin{pmatrix}
1 & 0 & 0 & 0 \\
0 & -1 & 0 & 0 \\
0 & 0 & 1 & 0 \\
0 & 0 & 0 & -1 \\
\end{pmatrix},
\eea where a positive (negative) eigenvalue of $K$ corresponds to
a right (left) mover state. In the earlier discussions, we first
re-express $\chi$ fields in terms of bosonic dual fields $\phi_l$
and $\theta_l$ (l=1,2) and then make a linear combination of
these new dual fields to define the bonding and anti-bonding
fields $\phi_{\pm}$ and $\theta_{\pm}$. Here we describe
the above process in a more compact way by introducing a set of
vectors ${l_i}$ ($i=1,2,3,4$): \bea
\Psi_1&=&\phi_+=l_1^T\chi=\frac{1}{2}(1,1,1,1)\chi \nonumber \\
\Psi_2&=&\theta_+=l_2^T\chi=\frac{1}{2}(1,-1,1,-1)\chi \nonumber \\
\Psi_3&=&\phi_-=l_3^T\chi=\frac{1}{2}(1,1,-1,-1)\chi \nonumber \\
\Psi_4&=&\theta_-=l_4^T\chi=\frac{1}{2}(1,-1,-1,1)\chi. \eea With these vectors, we can rewrite our Lagrangian under this new bases
$\Psi=(\Psi_1,\Psi_2,\Psi_3,\Psi_4)^T$. Define a transformation
matrix $U=(l_1^T,l_2^T,l_3^T,l_4^T)$ and we have $\Psi=U\chi$. The
new K matrix is \bea \tilde{K}=U^T K U=
\begin{pmatrix}
0 & 1 & 0 & 0 \\
1 & 0 & 0 & 0 \\
0 & 0 & 0 & 1 \\
0 & 0 & 1 & 0 \\
\end{pmatrix}.
\eea The block-diagonal form of $\tilde{K}$ indicates that the two
pairs of bosonic fields are completely decoupled. On one hand,
interaction will always gap out the bosonic anti-binding fields,
which corresponds to the lower half block of $\tilde{K}$. On the
other hand, when Luttinger parameter $K_+<1$ is satisfied,
$H_{\alpha_3}$ is suppressed and $U(1)_m$ symmetry emerges. In
this case, the bosonic bonding fields survive, and $\tilde{K}$
effectively reduces to a $2\times2$ matrix: \be
K_{eff}=\begin{pmatrix}
0 & 1 \\
1 & 0 \\
\end{pmatrix},
\ee which is consistent with the K matrix for a bosonic SPT phase
\cite{lu2012}.

\section{Conclusion}
In this paper, we have studied interacting topological phases in
thin films of a TMKI, focusing on
the bilayer system with the mirror Chern number $\pm2$. At the
single-particle level, we find that topological mirror insulator
phases with different mirror Chern numbers (from $\pm1$ to $\pm
6$) can be achieved by tuning film thickness and the hybridization
between different layers.
%It has been well established that quantum confinement can lead to oscillatory behavior between topological phases and trivial phases in topological insulators \cite{liu2010}. Besides this mechanism, we find strong hybridization between different layers can lead to additional band inversion in topological mirror Kondo insulators.
By introducing interaction into this system, bosonic SPT phases
can be realized under certain parameter regime in TMKI films.
Interaction can also drive the system into a
MSB phase, in which a domain wall between
different MSB order parameters can carry both
charge and spin. Current experimental studies of TMKIs are focusing on bulk materials, such as bulk SmB$_6$
\cite{neupane2013,jiang2013,xu2013,kim2014,zhang2013}, and we hope that
our studies on interacting topological phases can motivate more experimental explorations on TMKI thin films
\cite{yong2014}.

%Experimentally, we notice that SmB$_6$ thin films have been successfully grown by co-sputtering both SmB$_6$ and B targets . However, the thickness of the above system is about $100$ nm, and thus the thin film is still a bulk sample whose low energy physics is dominated by the surface states. To observe the 2D TCI phase we proposed, it is important to further reduce the thin film thickness to reach the true 2D limit. Additionally, substrates should be carefully grown in a symmetrical way to preserve the out-of-plane mirror symmetry of our systems. In the interacting limit, besides the existence of an exciting BSPT phase, we have shown that interesting physics happens even in the "trivial" regime. The emergence of domain wall excitations in a mirror symmetry breaking phase fractionalize the original bosonic degree of freedom into a fermionic excitation which carries a unit charge and a half-integer spin. Experimentally, short noise measurement has successfully verified the fundamental electric charge in both FQHE and superconductor systems \cite{saminadayar1997,reznikov1999,jehl2000}. We expect this detection approach will shed light on the bosonic nature of 2D TCI boundary modes as well as the corresponding fermionic domain wall excitations.

\section{Acknowledgement}
Rui-Xing Zhang would like to thank Jia-Bin Yu and Jian-Xiao Zhang
for helpful discussions. C.-X.L. acknowledges the support from
Office of Naval Research (Grant No. N00014-15-1-2675).

\bibliography{TKI}

\appendix

\section{Origin of $H_{\alpha_{3}}$}

In this section, we will show that $H_{\alpha_3}$ can originate from the standard Coulomb exchange interaction. This result not only validates the physical meaning of $H_{\alpha_3}$, but also implies that the Coulomb interaction generally breaks ``pseudo-spin" $U(1)_m$ symmetry, despite the fact that it preserves both spin $U(1)_s$ symmetry and mirror symmetry $M_z$.

To start with, let us first express a right-moving edge state fermion operator $\psi_{l,R}$ in terms of bulk fermion operators $c_{\alpha,\sigma}$,
\bea
\psi_{l,R}=\sum_{\alpha}f_{l,\alpha}c_{\alpha,\ua}+\sum_{\beta}g_{l,\beta}c_{\beta,\da},
\label{Eq:Right mover expansion}
\eea
 where $l=1,2$ and $c_{\alpha,\ua}$ and $c_{\beta,\da}$ labels annihilation operators for the bulk states while $f_{l,\alpha}$ and $g_{l,\beta}$ are for the envelope wave functions of edge modes. Here $\alpha$ and $\beta$ include both the orbital and layer indices for short. Let's label the states $|\alpha(\beta),\sigma\rangle=c^{\dag}_{\alpha(\beta),\sigma}|0\rangle$, where $|0\rangle$ is the vacuum state and $\sigma=\uparrow,\downarrow$. Since $\psi_{l,R}$ has a definite $M_z$ mirror parity $+i$ while $f_{l,\alpha}$ and $g_{l,\beta}$ have no spatial dependence along the z-direction, we require both the states $|\alpha,\uparrow\rangle$ and $|\beta,\downarrow\rangle$ possessing mirror parity $+i$. On the other hand, since
\bea
M_z
\begin{pmatrix}
|\ua\rangle \\
|\da\rangle \\
\end{pmatrix}
=
\begin{pmatrix}
+i & 0 \\
0 & -i \\
\end{pmatrix}
\begin{pmatrix}
|\ua\rangle \\
|\da\rangle \\
\end{pmatrix}
\eea
for spin parts, this indicates that $M_z|\alpha\rangle=+|\alpha\rangle$ and $M_z|\beta\rangle=-|\beta\rangle$.
%This means that for a state created by $c^{\dagger}_{\alpha,\ua}$, the mirror parity of its orbital part $\alpha$ is $+1$. While for a state created by $c^{\dagger}_{\beta,\da}$, the mirror parity of its orbital part $\beta$ is $-1$. The explicit forms of $\alpha$ and $\beta$ can be determined by Eq. (\ref{Eq:Mirror eigenstate}), which are actually not important in our present discussions.

TR symmetry transforms a right-mover $\psi_{l,R}$ to a left-mover $\psi_{l,L}$, giving rise to
\bea
\psi_{l,L}=\sum_{\alpha,\beta}f_{l,\alpha}^*c_{\alpha,\da}-g_{l,\beta}^*c_{\beta,\ua}.
\label{Eq:Left mover expansion}
\eea
It is easy to see that both $c_{\alpha,\da}$ and $c_{\beta,\ua}$ have mirror parity $-i$, which is consistent with that of $\psi_{l,L}$.

Now we are ready to rewrite the boundary interaction $H_{\alpha_{3}}$, as well as $H_{\alpha_{2,3}}$, in terms of the bulk fermionic operators $c_{\alpha/\beta,\sigma}$. For simplicity, we only consider one $\alpha$ state and one $\beta$ state in the decomposition of the edge state $\psi_{l,R/L}$. Let us first rewrite $H_{\alpha_3}$ with the help of Eq. (\ref{Eq:Right mover expansion}) and Eq. (\ref{Eq:Left mover expansion}) as
\begin{widetext}
\bea
H_{\alpha_3}&=&\alpha_3\sum_{k,k',q} \psi^{\dagger}_{1,L,k+q}\psi_{1,R,k}\psi_{2,L,k'-q}^{\dagger}\psi_{2,R,k'}+h.c. \nonumber \\
&=&\alpha_3\sum_{k,k',q}(f_{1,\alpha}c_{\alpha,\da,k+q}^{\dagger}-g_{1,\beta}c_{\beta,\ua,k+q}^{\dagger})(f_{1,\alpha}c_{\alpha,\ua,k}+g_{1,\beta}c_{\beta,\da,k}) \nonumber \\
&&\ \times(f_{2,\alpha}c_{\alpha,\da,k'-q}^{\dagger}-g_{2,\beta}c_{\beta,\ua,k'-q}^{\dagger})(f_{2,\alpha}c_{\alpha,\ua,k'}+g_{2,\beta}c_{\beta,\da,k'})+h.c. \nonumber \\
&=&\alpha_3\sum_{k,k',q}f_{1,\alpha}f_{2,\alpha}g_{1,\beta}g_{2,\beta}(c^{\dagger}_{\alpha,\da,k+q}c_{\alpha,\ua,k'}c^{\dagger}_{\beta,\ua,k'-q}c_{\beta,\da,k}
+c^{\dagger}_{\alpha,\da,k'-q}c_{\alpha,\ua,k}c^{\dagger}_{\beta,\ua,k+q}c_{\beta,\da,k'})\nonumber \\
&&\ -f_{1,\alpha}^2g_{2,\beta}^2c^{\dagger}_{\alpha,\da,k+q}c_{\alpha,\ua,k}c^{\dagger}_{\beta,\ua,k'-q}c_{\beta,\da,k'}
-g_{1,\beta}^2f_{2,\alpha}^2c^{\dagger}_{\alpha,\da,k'-q}c_{\alpha,\ua,k'}c^{\dagger}_{\beta,\ua,k+q}c_{\beta,\da,k}+h.c. \nonumber \\
&=&\alpha_3\sum_{k,k',q}[2f_{1,\alpha}f_{2,\alpha}g_{1,\beta}g_{2,\beta}-(f_{1,\alpha}^2g_{2,\beta}^2+g_{1,\beta}^2f_{2,\alpha}^2)]
c^{\dagger}_{\alpha,\da,k+q}c_{\alpha,\ua,k}c^{\dagger}_{\beta,\ua,k'-q}c_{\beta,\da,k'}+h.c.
\label{Eq:H_a3 Coulomb}
\eea
\end{widetext}

By defining the spin operator as
\bea
\hat{S^i}_{\alpha}(q)=\sum_{k}\sum_{\sigma,\sigma'}c^{\dagger}_{\alpha,\sigma,k+q}S^i_{\sigma,\sigma'}c_{\alpha,\sigma',k},
\eea
with $i\in\{x,y,z\}$ and $S^i$ is the corresponding Pauli matrix, $H_{\alpha_3}$ can be written in a compact form as
\bea
H_{\alpha_3}=J\sum_{q}\hat{S}^+_{\alpha}(q)\hat{S}^-_{\beta}(-q)+h.c.
\label{Eq:H_a3 exchange}
\eea
where
\bea
S^{\pm}&=&S^x\pm iS^y \nonumber
\eea
and
\bea
J&=&\alpha_3[2f_{1,\alpha}f_{2,\alpha}g_{1,\beta}g_{2,\beta}-(f_{1,\alpha}^2g_{2,\beta}^2+g_{1,\beta}^2f_{2,\alpha}^2)]. \nonumber \\
\eea
Eq. (\ref{Eq:H_a3 exchange}) is exactly the Coulomb exchange interaction for the multi-orbital case. We would like to emphasize that this exchange process occurs between two electrons with opposite $M_z$ mirror parities for the $\alpha$ and $\beta$ parts in their wave functions. According to Eq. (\ref{Eq:H_a3 Coulomb}), this type of Coulomb interaction breaks $U(1)_m$ symmetry, while preserving both $U(1)_s$ and $M_z$ symmetry.

\end{document}